\documentclass[aps,preprint]{revtex4}
\usepackage{amsmath}
\usepackage{mathrsfs}
\usepackage{bm}

\begin{document}

\title{Nonlinear interactions between kinetic Alfv\'{e}n and ion-sound waves}
\author{G. Brodin, L. Stenflo and P. K. Shukla \\
Department of Physics, Ume\aa\ University, SE-901 87 Ume\aa , Sweden}

\begin{abstract}
The resonant interaction between kinetic Alfv\'{e}n and ion-acoustic waves
is considered using the Hall-MHD theory. The results of previous authors are
generalized to cover both finite Larmor radius as well as the ideal MHD
results. It is found that the three wave coupling is strongest when the
wavelength is comparable to the ion-sound gyroradius. Applications of our
work to weak turbulence theories as well as to the heating of the solar
corona are pointed out.
\end{abstract}

\maketitle


The nonlinear interaction of magnetohydrodynamic (MHD) waves has been
considered by numerous authors (see for example, Sagdeev and Galeev, 1969;
Petviashvili and Pokhotelov, 1992; Shukla and Stenflo, 1999). The
applications involve fusion plasmas (Hasegawa and Uberoi, 1982), space
physics (Petviashvili and Pokhotelov, 1992; Shukla and Stenflo, 1999; Wu and
Chao, 2004) as well as solar physics (Shukla et al., 1999; Voitenko and
Goossens, 2000 and 2002; Shukla and Stenflo, 2005; Chandran 2005) and
astrophysics (Ng and Bhattacharjee, 1996; Goldreich and Sridhar, 1997). The
classic work on three wave interaction of ideal MHD waves (Sagdeev and
Galeev, 1969) was later generalized to account for arbitrary (but still
ideal) MHD wave modes and directions of propagations (Brodin and Stenflo,
1988). The ideal MHD processes were soon suggested to have applications for
the heating of fusion plasmas (Lashmore-Davies and Ong, 1974). Hasegawa and
Chen (1976a) showed, however, that processes involving kinetic Alfv\'{e}n
waves were more efficient for that purpose. The latter waves can be
described by the Hall-MHD theory, and general three wave coupling
coefficients for the Hall-MHD plasmas were thus deduced by Brodin and
Stenflo (1990). Applications for the parametric decay instability of
magneto-acoustic waves into two kinetic Alfv\'{e}n waves, to the heating of
the solar corona, were considered by Voitenko and Goossens (2002). The Joule
electron heating caused by high-frequency dispersive Alfv\'{e}n waves in the
solar corona was also analysed by Shukla et al. (1999).

Much of the previous work describing parametric instabilities involving
kinetic Alfv\'{e}n waves (KAWs) has adopted a kinetic theory (Hasegawa and
Chen, 1976; Voitenko 1998) or multi-fluid models (Erokhin, Moiseev and
Mukhin, 1978; Voitenko and Goossens, 2002). In the present paper, we will
however demonstrate that the essential characteristics of the three-wave
decay interaction involving the KAWs can be more simply described within a
unified formalism of the Hall-MHD theory. An important result of that
formalism is that the decay of kinetic Alfv\'{e}n waves is dominated by the
excitation of modes with short perpendicular wavelengths, of the order of
the ion-sound gyroradius, that must be described by the Hall-MHD theory. We
shall show that this specific example has general significance, and that the
ideal MHD typically is unable to deal with the nonlinear evolution of the
MHD waves, even if the initial conditions are within the range of the ideal
MHD.

Thus, we start with the Hall-MHD equations, that can be written as

\begin{equation}
\frac{\partial \rho }{\partial t}+\nabla \cdot (\rho \mathbf{v})=0,
\end{equation}
\begin{equation}
\rho \frac{d\mathbf{v}}{dt}=-c_{s}^{2}\nabla \rho +\frac{(\nabla \times 
\mathbf{B})\times \mathbf{B}}{\mu _{0}},
\end{equation}
and 
\begin{equation}
\frac{\partial \mathbf{B}}{\partial t}=\nabla \times (\mathbf{v}\times 
\mathbf{B}-\frac{m_{i}}{e}\frac{d\mathbf{v}}{dt}),
\end{equation}
where $d/dt=\partial /\partial t+\mathbf{v}\cdot \nabla $ , $e$ \ and $m_{i}$
is the ion charge and mass, whereas $\rho $, $\mathbf{v}$, and $\mathbf{B}$
are the density, velocity and magnetic field, respectively, and $%
c_{s}=[\left( T_{e}+T_{i}\right) /m_{i}]^{1/2}$ is the ion-sound speed. Here 
$T_{e}$ and $T_{i}$ are the electron and ion temperatures.

Considering the resonant interaction between three waves which satisfy the
matching conditions

\begin{equation}
\omega _{3}=\omega _{1}+\omega _{2},
\end{equation}
and 
\begin{equation}
\mathbf{k}_{3}=\mathbf{k}_{1}+\mathbf{k}_{2},
\end{equation}
we can, using (1)-(5), derive the equations [see Brodin and Stenflo (1990)
for details]

\begin{equation}
\left( \frac{\partial }{\partial t}+\mathbf{v}_{g1,2}\cdot \nabla \right)
\rho _{1,2}=-\frac{1}{\partial \tilde{D}_{1,2}/\partial \omega _{1,2}}C\rho
_{2,1}^{\ast }\rho _{3},  \label{Coupling1}
\end{equation}
and 
\begin{equation}
\left( \frac{\partial }{\partial t}+\mathbf{v}_{g3}\cdot \nabla \right) \rho
_{3}=\frac{1}{\partial \tilde{D}_{3}/\partial \omega _{3}}C\rho _{1}\rho
_{2},  \label{Coupling2}
\end{equation}
where 
\begin{eqnarray}
C &=&\frac{\omega _{1}\omega _{2}\omega _{3}}{\rho _{0}k_{1\perp
}^{2}k_{2\perp }^{2}k_{3\perp }^{2}}\left[ \frac{\mathbf{K}_{3}\cdot \mathbf{%
K}_{2}^{\ast }}{\omega _{1}}k_{1\perp }^{2}+\frac{\mathbf{K}_{3}\cdot 
\mathbf{K}_{1}^{\ast }}{\omega _{2}}k_{2\perp }^{2}+\frac{\mathbf{K}%
_{1}^{\ast }\cdot \mathbf{K}_{2}^{\ast }}{\omega _{3}}k_{3\perp }^{2}\right.
-  \notag \\
&&\frac{k_{1\perp }^{2}k_{2\perp }^{2}k_{3\perp }^{2}}{\omega _{1}\omega
_{2}\omega _{3}}c_{s}^{2}+\frac{i\omega _{ci}}{\omega _{3}}(\frac{k_{2z}}{%
\omega _{2}}-\frac{k_{1z}}{\omega _{1}})\left( (\mathbf{K}_{3}+\frac{i\omega
_{3}\mathbf{k}_{3}\times \mathbf{K}_{3}}{\omega _{ci}k_{3z}})\cdot (\mathbf{K%
}_{1}^{\ast }-\right.  \notag \\
&&\left. \left. \frac{i\omega _{1}\mathbf{k}_{1}\times \mathbf{K}_{1}^{\ast }%
}{\omega _{ci}k_{1z}})\times (\mathbf{K}_{2}^{\ast }-\frac{i\omega _{2}%
\mathbf{k}_{2}\times \mathbf{K}_{2}^{\ast }}{\omega _{ci}k_{2z}})-\mathbf{K}%
_{3}\cdot (\mathbf{K}_{1}^{\ast }\times \mathbf{K}_{2}^{\ast })\right) %
\right] ,
\end{eqnarray}
\begin{eqnarray}
\tilde{D}_{j} &=&\Biggl[\omega _{j}^{4}-\omega
_{j}^{2}k_{j}^{2}(c_{A}^{2}+c_{s}^{2})+k_{jz}^{2}k_{j}^{2}c_{A}^{2}c_{s}^{2}
\notag \\
&&-\frac{\omega _{j}^{2}k_{jz}^{2}k_{j}^{2}(\omega
_{j}^{2}-k_{j}^{2}c_{s}^{2})c_{A}^{4}}{\omega _{ci}^{2}(\omega
_{j}^{2}-k_{jz}^{2}c_{A}^{2})}\Biggr]\frac{(\omega
_{j}^{2}-k_{j}^{2}c_{s}^{2})}{\omega _{j}^{2}k_{j\perp
}^{2}k_{j}^{2}c_{A}^{2}},  \label{Diespersionfull}
\end{eqnarray}
and 
\begin{equation}
\mathbf{K}_{j}=\mathbf{k}_{j\perp }\frac{(\omega
_{j}^{2}-k_{jz}^{2}c_{s}^{2})}{\omega _{j}^{2}}+\frac{i\hat{\mathbf{z}}%
\times \mathbf{k}_{j\perp }(\omega
_{j}^{2}-k_{j}^{2}c_{s}^{2})k_{jz}^{2}c_{A}^{2}}{\omega _{ci}\omega
_{j}(\omega _{j}^{2}-k_{jz}^{2}c_{A}^{2})}+\frac{k_{j\perp
}^{2}k_{jz}c_{s}^{2}}{\omega _{j}^{2}}\hat{\mathbf{z}}.  \label{Full-K}
\end{equation}
Here $\mathbf{v}_{gj}$ is the group velocity of wave $j$, $\omega _{ci}$ is
the ion gyrofrequency, and $c_{A}=(B_{0}/\mu _{0}\rho _{0})^{1/2}$ is the
Alfv\'{e}n speed. The derivation of (\ref{Coupling1}) and (\ref{Coupling2})
is straightforward (Brodin and Stenflo, 1990). Our result has the
significant advantage that the same coupling coefficient $C$ appears in both
(\ref{Coupling1}) and (\ref{Coupling2}). This means that the Manley-Rowe
relations are always satisfied. We could alternatively have used, instead of 
$\rho _{j}$, longitudinal and/or transverse components of the velocity
(using the relation $\mathbf{v}_{j}=\rho _{j}\mathbf{K}_{j}\omega
_{j}/k_{j\perp }^{2}\rho _{0}$), where the transverse velocity component is
particularly convenient for Alfv\'{e}n waves with small or vanishing density
perturbations.

Here, we focus on wave modes with frequencies well below the ion
gyrofrequency, but with large perpendicular wavenumbers, so that $k_{\perp
}^{2}c_{s}^{2}/\omega _{ci}^{2}$ can be of order unity. For the particular
case of the KAWs, and for an intermediate beta plasma with $%
(m_{e}/m_{i})c_{A}^{2}<c_{s}^{2}\ll c_{A}^{2}$, where $m_{e}$ is the
electron mass, (\ref{Full-K}) can then be approximated by

\begin{equation}
\mathbf{K}\approx -i\frac{\omega _{ci}}{\omega }\hat{\mathbf{z}}\times 
\mathbf{k+}\frac{k_{\perp }^{2}k_{z}c_{s}^{2}}{\omega ^{2}}\hat{\mathbf{z}},
\end{equation}
(where the second term is smaller than the first by a factor of the order $%
c_{s}/c_{A}$), together with the approximate dispersion relation

\begin{equation}
\omega ^{2}=k_{z}^{2}c_{A}^{2}\left[ 1+\frac{k_{\perp }^{2}c_{s}^{2}}{\omega
_{ci}^{2}}\right].  \label{KAW-disp}
\end{equation}
Similarly, for the the ion-acoustic waves we can write

\begin{equation}
\mathbf{K}\approx \frac{ik_{\perp }^{2}c_{s}^{2}\hat{\mathbf{z}}\times 
\mathbf{k}}{\omega _{ci}\omega }+\frac{k_{\perp }^{2}k_{z}c_{s}^{2}}{\omega
^{2}}\hat{\mathbf{z}},
\end{equation}
provided that $\omega ^{2}/\omega _{ci}^{2}\ll k_{z}/k_{\perp }$. The
corresponding dispersion relation can then be approximated as

\begin{equation}
\omega ^{2}=\frac{k_{z}^{2}c_{s}^{2}}{(1+k_{2\perp }^{2}c_{s}^{2}/\omega
_{ci}^{2})}.  \label{IA-disp}
\end{equation}

Next, considering two waves \ (with index 1 and 3) to be kinetic Alfv\'{e}n
waves described by (\ref{KAW-disp}), \ and one wave (with index 2) to be an
ion-acoustic wave, described by (\ref{IA-disp}), the interaction equations
can be rewritten as

\begin{equation}
\left( \frac{\partial }{\partial t}+\mathbf{v}_{g1}\cdot \nabla \right)
v_{1}=-\frac{i\omega _{1}C_{AmA}\rho _{2}^{\ast }v_{3}}{\rho _{0}},
\end{equation}
\begin{equation}
\left( \frac{\partial }{\partial t}+\mathbf{v}_{g2}\cdot \nabla \right) \rho
_{2}=i\frac{\rho _{0}k_{2z}^{2}}{\omega _{2}}C_{AmA}v_{1}^{\ast }v_{3},
\end{equation}
and 
\begin{equation}
\left( \frac{\partial }{\partial t}+\mathbf{v}_{g3}\cdot \nabla \right)
v_{3}=-\frac{i\omega _{3}C_{AmA}v_{1}\rho _{2}}{\rho _{0}},
\end{equation}
with the coupling coefficient approximated by

\begin{equation}
C_{AmA}=\cos \theta -\frac{\omega _{2}^{2}k_{1\perp }k_{3\perp }}{%
k_{2}^{2}c_{A}^{2}k_{1z}k_{3z}}\sin ^{2}\theta -\mathrm{i}\frac{%
c_{s}^{2}\sin \theta }{k_{2z}\omega _{ci}}\left( \frac{k_{3z}}{\omega _{3}}-%
\frac{k_{1z}}{\omega _{1}}\right) \left[ k_{1\perp }^{2}+k_{3\perp
}^{2}-k_{2\perp }^{2}+\frac{k_{1\perp }^{2}k_{3\perp }^{2}c_{s}^{2}}{\omega
_{ci}^{2}}\right]  \label{Coeff2}
\end{equation}
where $v_{1,3}$ is the magnitude of the velocity of waves 1 and 3,
respectively, and $\theta $ is the angle between $\mathbf{k}_{1\perp }$ and $%
\mathbf{k}_{3\perp }$ (or the angle between $\mathbf{v}_{1}$ and $\mathbf{v}%
_{3}$ when $\mathbf{k}_{1,3\perp }\rightarrow 0$). The first two terms in (%
\ref{Coeff2}) dominate for $k_{j\perp }^{2}c_{s}^{2}\ll \omega _{ci}^{2}$,
and agree with the ideal MHD coupling coefficient of Brodin and Stenflo
(1988) in the low-beta limit considered here. The third term, which
dominates for large perpendicular wavenumbers, agrees with the coupling
coefficient of Hasegawa and Chen (1976), which was derived using a kinetic
approach. As a specific example, we let wave 3 be a pump wave. To
demonstrate the importance of the second term in (\ref{Coeff2}), we assume
that all waves have large perpendicular wavenumbers, such that $k_{\perp
}^{2}c_{s}^{2}/\omega _{ci}^{2}\sim 1$. Furthermore, to facilitate an order
of magnitude estimate of (\ref{Coeff2}) we let $\mathbf{k}_{1\perp }$ and $%
\mathbf{k}_{3\perp }$ be approximately perpendicular to each other. In this
case, the magnitude of the third part of $C_{AmA}$ can be estimated as

\begin{equation}
C_{AmA}\sim \frac{\omega _{ci}}{\omega _{3}}\gg 1,  \label{coeff-estimate}
\end{equation}
which is much larger than the first two parts of $C_{AmA}$ accounted for by
the ideal MHD, and which do not exceed unity. As a consequence, the growth
rate $\Gamma _{\mathrm{HM}}$ into short perpendicular wavelength modes (with 
$k_{\perp }^{2}c_{s}^{2}/\omega _{ci}^{2}\sim 1$, described by the Hall-MHD
theory) is larger than the growth rate $\Gamma _{\mathrm{IM}}$ due to the
ordinary ideal MHD modes by a factor $\Gamma _{\mathrm{HM}}/\Gamma _{\mathrm{%
IM}}\sim \omega _{ci}/\omega $. Thus, the increased coupling strength into
short wavelength modes with perpendicular wavelengths of the order of the
ion-sound gyroradius ($c_{s}/\omega _{ci}$) affects the parametric decay
processes significantly. This is very important as the wave cascade
processes (Goldreich and Sridhar, 1997) of weak turbulence theories are
based on the resonant three-wave coupling mechanism. Moreover, while the
general features of such processes lead to a broadening of the frequency
spectrum, and energy transfer towards lower frequencies, we note that the
energy transfer will mainly occur in the direction of higher coupling
strength, i.e. into modes with short perpendicular wavelengths. Thus, even
for an initial turbulent spectrum well within the range of the ideal MHD,
wave cascade processes will eventually lead to the excitation of short
perpendicular wavelengths and the necessity to use the Hall-MHD rather than
the ideal MHD.

Assuming that wave 3 is a pump wave with magnetic field magnitude $B_{3}$,
and using the estimate $B_{3}\sim v_{3}B_{0}/c_{A}$, we find the growth rate 
\begin{equation}
\Gamma _{\mathrm{HM}}\sim \omega _{ci}\frac{B_{3}}{B_{0}}
\label{Growth-estimate}
\end{equation}
It should, however, be pointed out that the present decay channel for a KAW
into an ion-acoustic wave and another KAW is not unique. Other decay
channels that have been investigated for KAWs can compete with it (e.g.
Voitenko and Goossens 2000; Onishchenko et. al. 2004). \ These processes can
spread out the KAW spectrum and thus prevent the parametric decay into
ion-acoustic waves. To find out the relative importance of the decay into
ion-acoustic waves as compared to the above mentioned processes, we should
therefore compare our estimate (\ref{Growth-estimate}) with the growth rates 
$\Gamma _{AA}$ of Voitenko and Goossens (2000), and $\Gamma _{\mathrm{JGR}}$
of Onishchenko et al. (2004). We then use the estimates $\Gamma _{AA}\sim
0.2\omega _{3}k_{3}^{2}\rho _{i}B_{3}/k_{3z}B_{0}$, where $\rho _{i}=\left(
T_{i}/m_{i}\right) ^{1/2}/\omega _{ci}$ is the ion Larmor radius, and $%
\Gamma _{\mathrm{JGR}}\sim 2\omega
_{3}Dk_{3}^{2}B_{3}^{2}/k_{3z}^{2}B_{0}^{2}$, where $D$ is a factor of order
unity (Onishchenko et al., 2004). Onishchenko et al. (2004) showed that $%
\Gamma _{\mathrm{JGR}}$ is smaller than $\Gamma _{AA}$ if $B_{3}/B_{0}$ is
smaller than a factor of the order $k_{3z}\rho _{i}$. A comparison between $%
\Gamma _{\mathrm{HM}}$ of the present paper and $\Gamma _{AA}$ reveals that $%
\Gamma _{\mathrm{HM}}/\Gamma _{AA}\sim 5\omega _{ci}k_{3z}/\omega
_{3}k_{3}^{2}\rho _{i}$. Although the estimates above are very crude, they
show that the process we consider in the present paper can be even more
important than those of previous papers for a significant range of
parameters.

To summarize, we have reconsidered the interaction of kinetic Alfv\'{e}n and
ion-acoustic waves using the Hall-MHD theory. In particular, the three wave
equations involving the nonlinear coupling between two kinetic Alfv\'{e}n
waves and one ion-acoustic wave have been explicitly presented. The same
coupling coefficient (\ref{Coeff2}) appears in all these equations, implying
that the Manley-Rowe relations are fulfilled. Furthermore, our coupling
coefficient (\ref{Coeff2}) includes both the ideal MHD results of Brodin and
Stenflo (1988), and the effects due to the kinetic approach of Hasegawa and
Chen (1976a), in a unified formalism. As can be seen from (\ref{Coeff2}),
the wave coupling is strongest for perpendicular wavelengths of the order of
the ion-sound gyroradius. As has been argued above, this has important
consequences for several processes, such as for the parametric decay
instabilities and wave cascades in weak turbulence theories. Moreover, the
formalism presented above is relevant for plasma particle energization in
the solar corona by kinetic Alfv\'{e}n waves. In the solar corona, a kinetic
Alfv\'{e}n pump wave can be excited by a linear transformation of an Alfv%
\'{e}n surface wave in the neighbourhood of the resonance region (Hasegawa
and Chen, 1976b). The mode converted kinetic Alfv\'{e}n wave can then
further decay into a daughter kinetic Alfv\'{e}n wave and a dispersive ion
sound wave, as described here. The nonlinearly excited kinetic Alfv\'{e}n
waves can attain large amplitudes and small perpendicular wavelengths
(Hasegawa and Chen, 1976b), and they could therefore be our most efficient
agents for  energization of ions and electrons by kinetic Alfv\'{e}n wave
phase mixing and Joule heating (Ionson, 1978; Hasegawa and Uberoi, 1982;
Shukla et al., 1994; Cramer, 2001), as well as for turbulent heating and
particle-KAW interactions.

This research was partially supported by the Swedish Research Council.

\end{document}